\documentclass[aps,pra,reprint]{revtex4-1}
\usepackage{graphicx}
\usepackage{amsmath,amsthm,amssymb,latexsym,amsfonts,times,mathrsfs,verbatim, lipsum}
\usepackage{bm}
\usepackage{color}

\begin{document}

\title{Coherent Magneto-Optomechanical Signal Transduction and Long-Distance Phase-Shift Keying}

\author{M.J. Rudd}
\author{P.H. Kim}
\author{C.A. Potts}
\author{C. Doolin}
\author{H. Ramp}
\author{B.D. Hauer}
\author{J.P. Davis}\email{jdavis@ualberta.ca}
\address{Department of Physics, University of Alberta, Edmonton, Alberta, Canada T6G 2E9}

\begin{abstract}
A transducer capable of converting quantum information stored as microwaves into telecom-wavelength signals is a critical piece of future quantum technology as it promises to enable the networking of quantum processors.  Cavity optomechanical devices that are simultaneously coupled to microwave fields and optical resonances are being pursued in this regard. Yet even in the classical regime, developing optical modulators based on cavity optomechanics could provide lower power or higher bandwidth alternatives to current technology.  Here we demonstrate a magnetically-mediated wavelength conversion technique, based on mixing high frequency tones with an optomechanical torsional resonator.  This process can act either as an optical phase or amplitude modulator depending on the experimental configuration, and the carrier modulation is always coherent with the input tone.  Such coherence allows classical information transduction and transmission via the technique of phase-shift keying.  We demonstrate that we can encode up to eight bins of information, corresponding to three bits, simultaneously and demonstrate the transmission of an 52,500 pixel image over 6 km of optical fiber with just 0.67\% error.  Furthermore, we show that magneto-optomechanical transduction can be described in a fully quantum manner, implying that this is a viable approach to signal transduction at the single quantum level.

\end{abstract}

\maketitle

\section{\label{sec:level1}Introduction}

The ability to reliably convert microwave photons to telecom-wavelength photons promises to enable long-distance entanglement of superconducting qubits over telecom fiber networks, a major advance towards developing a quantum internet \cite{Kimble08}.  Cavity optomechanics, in which mechanical and optical (or microwave) degrees of freedom are highly coupled, is a leading contender for such wavelength conversion, in part because mechanical motion can be coupled to a wide range of cavities along the electromagnetic spectrum \cite{Hil12,Lecocq17}. Despite such promise, there are many obstacles to efficient and low-noise optomechanical wavelength conversion, such as parasitic thermal noise \cite{Ramp18,Forsch18} and the difficulty of simultaneously coupling two widely separated cavity resonances, such as microwave and telecom.  The later stems primarily from the natural mismatch between the size and materials of mechanical resonators that couple to microwaves versus optical wavelengths.  Nonetheless, a number of clever approaches are being pursued to circumvent this mismatch, such as dielectric membranes that are simultaneously coupled to a high-finesse Fabry-P\'erot cavity and patterned with a superconducting electrode for integration into a microwave cavity \cite{Andrews14}, and piezoelectric optomechanical resonators in which microwaves can be coupled to the mechanics either directly through an electric field \cite{Bochmann13,Witmer17} or indirectly via surface acoustic waves \cite{Balram16,Vainsencher16}.  

We have recently developed another option, using a magnetic component affixed to a telecom-wavelength optomechanical resonator \cite{Kim17}, to convert 300 MHz to 1.1 GHz (UHF) tones to a telecom carrier.  This takes advantage of a spin-mode intermediary \cite{Kim18}, not unlike the surface acoustic wave intermediary in piezoelectric conversion, but with increased utility -- as described below.  Here, we show both theoretically and experimentally that this conversion results in a telecom modulation that is coherent with the phase of the microwave drive tone.  We therefore use phase-shift keying (PSK) \cite{Anderson} to encode information in the microwave tone -- in up to 8 bins, and therefore, three bits simultaneously -- and read it out through the telecom-wavelength carrier.  Furthermore, we are able to use PSK to send an image originating in UHF tones, along a telecom fiber over a distance of 6 km with low ($\leq1$\%) error rate, demonstrating the potential for this wavelength conversion scheme to be integrated with long-distance fiber networks.    

\section{Experiment}

At the heart of our conversion technique is the realization that magnetic materials can be harnessed to interface with microwaves as they couple to magnetic fields \cite{Bai17}, just as piezoelectric materials couple to electric fields \cite{Bochmann13,Balram16}.  Therefore, a hard magnetic material (high magnetic coercivity) deposited onto a mechanical resonator allows magnetic control over its motion, as we have previously demonstrated by driving and damping an optomechanical torsional resonator \cite{Kim17}. In order to take advantage of the quality-factor enhancement of a mechanical resonator, such control is typically limited to a narrow band of frequencies around the mechanical resonance.  To extend the range of frequencies over which one can control the mechanical motion, and hence use for signal transduction, we instead use a soft magnetic material (low magnetic coercivity) with low-damping collective spin excitations.  In essence this allows one to oscillate the magnetic moment of the system, and then use the mixing inherent to magnetic torque to frequency-mix to the mechanical resonance \cite{Los15}.  

Here we use a Ni$_{80}$Fe$_{20}$ (permalloy) disk, which is patterned onto the mechanical resonator during nanofabrication.  In such a magnetic material the spin excitations have resonances at $f_\textrm{res}$ with susceptibility \cite{Kalarickal06},
\begin{equation}
\chi(f_x) = \frac{\chi_0 f_\textrm{res}^2}{\left[f_\textrm{res}^2-f_x\left(f_x-i\Delta f\right)\right]},
\end{equation}
which can be driven by an alternating magnetic field in the plane of the device ($\hat{x}$) with strength $H_x$ and frequency $f_x$.  Here $\Delta f$ is the spin-resonance linewidth and $\chi_0$ is the off-resonant susceptibility \cite{Kalarickal06}. The spin resonance used in the present experiment is shown in Fig.~1, and corresponds to the gyrotropic motion of a vortex state \cite{Park03,Novosad05} at $\sim$385 MHz. A second orthogonal magnetic field, with strength $H_z$ and frequency $f_z$, can then be applied to down-mix the induced magnetic moment to a torque along the torsion rod ($\hat{y}$), at the difference between the two magnetic drive frequencies \cite{Los15,Kim18}: 
\begin{equation}
\tau_y(|f_\textrm{x}-f_\textrm{z}|) =  \frac{V|H_x||H_z|\mu_0}{2}|\chi(f_\textrm{res})|,
\end{equation}
where $V$ is the volume of magnetic material and we consider the magnetic susceptibility to be driven on resonance.  To take advantage of the quality-factor enhancement of the resonator, the two alternating magnetic fields should be separated in frequency by the mechanical resonance frequency, \textit{i.e.}~$f_\textrm{m}=|f_x-f_z|$ \cite{Los15}.  In this scenario, the out-of-plane field can be considered the tone one wishes to coherently convert to telecom wavelengths, with the in-plane field as a fixed pump.  As the frequency of the collective spin modes can be controlled by the geometry \cite{Novosad05,Kim18} and are rather broadband ($\Delta f\sim30\,$MHz in Fig.~1), this torque-mixing provides a tunable and broadband route for signal transduction.  

\begin{figure}[t]
\includegraphics[width= 0.49\textwidth]{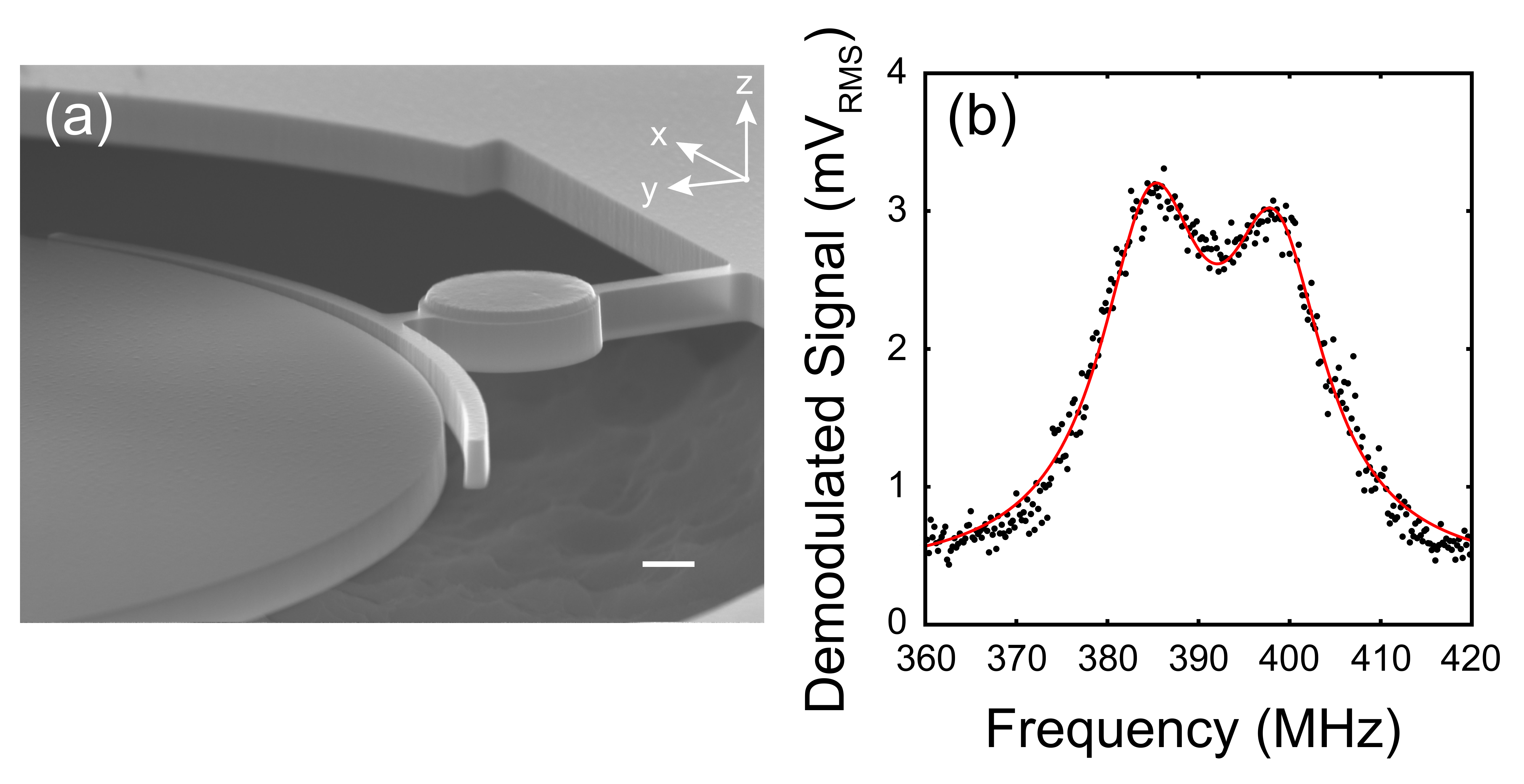}
\caption{a) Scanning electron micrograph of the magneto-optomechanical device.   Electron-beam lithography was used to pattern the device in 250-nm-thick single-crystal-silicon on silicon dioxide.  A second electron-beam-lithography step was used to deposit a 910 nm diameter, 60 nm thick, Ni$_{80}$Fe$_{20}$ disk onto the sample pad of the torsional resonator.  Scale bar is 200 nm.  (b)  Spin resonances in the Ni$_{80}$Fe$_{20}$ disk can be driven by modulating a magnetic field along the x-axis.  This results in an alternating in-plane magnetic moment in the disk, which -- in the presence of a second magnetic field along the z-axis -- produces a torque along the torsion axis of the device (y-axis) as described in the text.  The frequencies of the spin resonance and torsional resonance set the frequencies that can be used for the phase modulation presented below.  The origin of the doublet in the spin resonance is not currently understood and beyond the scope of the current work.}
\end{figure}

The optomechanical device used here is similar to those in Ref.~\citenum{Kim18}, specifically designed to be highly sensitive to torque \cite{Kim13,Kim16}.  It consists of a disk-shaped optical resonator that is partially encircled by a concentric mechanical resonator (Fig.\,1b), which is separated from the microdisk by a 100 nm gap.  The mechanical motion has a torsional resonance with a center frequency of $f_\textrm{m}=16.7$ MHz and a quality factor of 9,400 in vacuum at room temperature.  The mechanical displacement is read-out by monitoring the optical cavity transmission via a dimpled-tapered fiber \cite{Michael07,Hau14}.  This sensitive detection scheme results in a thermomechanically calibrated \cite{Hauer13} mechanical noise floor of 1.1 fm$/\sqrt\textrm{Hz}$, and a torque sensitivity of 0.4 zNm$/\sqrt\textrm{Hz}$.  A 910 nm diameter permalloy disk was deposited in ultra-high vacuum onto the ``sample pad'' of the torsional resonator, Fig.\,1, in a second e-beam lithography step, enabling the torque-mixing described above.  

High-frequency magnetic fields are applied by driving single-loop coils patterned on a printed circuit board (PCB) \cite{Kim17,Los15}.  Two orthogonal field directions are achieved by having the loop for the out-of-plane field printed on the top of the PCB, with the in-plane field generated by vias through the thickness of the PCB \cite{Kim17,Los15}.  These are driven using two signal generators, each referenced to a 10 MHz Rb standard.  In the current experiment, each is driven at 16 dBm.  A static magnetic field is also applied in the plane of the magnetic disk, to shift the vortex location and hence optimize the spin-resonance signal \cite{Los15,Kim18}.   

\section{Theory}

To investigate the signal transduction in our magneto-optomechanical architecture, we begin by considering the canonical optomechanical Hamiltonian describing coherent evolution of the system \cite{Purdy17},
\begin{equation}
\mathcal{H}_0 = \hbar\omega_\textrm{c}a^{\dagger}a + \frac{p^2}{2m} + \frac{m\Omega^2_\textrm{m}x^2}{2}+\hbar gxa^{\dagger}a.
\end{equation}
Here, $a^{\dagger}$ ($a$) is the creation (annihilation) operator for the optical cavity mode, $\omega_\textrm{c}$ is the resonance frequency of the optical cavity, $x$ and $p$ are the mechanical resonator position and momentum operators, $\Omega_\textrm{m} = 2\pi f_\textrm{m}$ is the mechanical resonance frequency, m is the effective mass of the mechanical element, and $g$ is the optomechanical coupling constant defined as $g = d\omega_\textrm{c} / dx$.

From the Hamiltonian we can derive the quantum Langevin equations \cite{Aspelmeyer14}.  We include terms describing the coupling between the system, the magnetic drive, and the external environment:
\begin{equation}
\dot{x} = \frac{p}{m},
\end{equation}
\begin{equation}
\dot{p} = -m\Omega^2_\textrm{m}x - \hbar ga^{\dagger}a-\Gamma_\textrm{m}p+\delta F_\textrm{th}+\delta F_\textrm{tor},
\end{equation}
and
\begin{equation}
\dot{a} = -i\omega_ca - igxa-\frac{\kappa}{2}a+\sqrt{\kappa_\textrm{e}}a_\textrm{in}.
\end{equation}
Specifically, $\delta F_\textrm{tor} = \tau_y / l_\textrm{eff}$ is the force applied via the torque mixing process, $l_\textrm{eff}$ is the effective length of the lever arm \cite{Kim17}, $\delta F_\textrm{th}$ is the random thermal forces driving the mechanical element, $\Gamma_\textrm{m}$ is the mechanical damping rate, $\kappa$ is the optical cavity damping rate, and $\kappa_\textrm{e}$ is the coupling rate to the external cavity drive.

Linearizing the equations of motion around the steady state optical amplitude $\bar{a}$, and solving them in the Fourier domain (see Appendix A), the final equations of motion are given by 
\begin{equation}
x[\omega] \approx \chi_\textrm{m}[\omega]
\Big( \delta F_\textrm{th} + \delta F_\textrm{tor} \Big)
\end{equation}
and
\begin{equation}
    \delta a[\omega] = \chi_\textrm{c}[\omega] \Big( -ig\bar{a}x[\omega]+\sqrt{\kappa_e}\delta a_\textrm{in}[\omega]
    \Big).
\end{equation}
In these equations $\chi_\textrm{m}[\omega] =(m(\Omega^2_\textrm{m}-\omega^2-i\Gamma_\textrm{m}\omega))^{-1}$ is the mechanical susceptibility, $\chi_\textrm{c}[\omega] = (\kappa/2 - i(\omega-\Delta_\textrm{l}))^{-1}$ is the optical cavity susceptibility, $\Delta_\textrm{l}$ is the laser detuning from the optical cavity resonance, and we have ignored effects from dynamical backaction \cite{Aspelmeyer14}. For a sufficiently strong torque, \textit{i.e.}~greater than the thermal force, the mechanical motion follows the applied torque directly (in-phase). 

Next, we consider the effect this driving force has on the detected optical quadrature operators (see Appendix B for definition).  Let us first consider optical homodyne detection of the phase quadrature. In this situation the laser is tuned directly on cavity resonance (\textit{i.e.}, $\Delta_\textrm{l} =0$) to maximize the optical signal. Solving for the output phase-quadrature we find,
\begin{eqnarray}
    \delta X_{Q,\textrm{out}}&&[\omega] = (1-\kappa_\textrm{e}\chi_c[\omega])\delta X_{Q,\textrm{in}}[\omega]\nonumber\\
    &&+2\sqrt{\kappa_\textrm{e}}\chi_c[\omega]\chi_\textrm{m}[\omega]g\bar{a}(\delta F_\textrm{th}+\delta F_\textrm{tor}).
\end{eqnarray}
In contrast, for direct detection the laser is tuned to the highest slope of the optical resonance (\textit{i.e.}, $\Delta_\textrm{l} = \kappa/2$). Direct detection is analogous to a measurement of the output amplitude-quadrature; therefore, solving for the amplitude operator we find,
\begin{eqnarray}
\delta X_{I,\textrm{out}}&&[\omega] = (1-\kappa_\textrm{e}\Re(\chi_\textrm{c}[\omega]))\delta X_{I,\textrm{in}}[\omega]\nonumber\\
&&+\kappa_\textrm{e}\Im(\chi_\textrm{c}[\omega])\delta X_{Q,\textrm{in}}[\omega]\\
&&-2\sqrt{\kappa_\textrm{e}}\Im(\chi_\textrm{c}[\omega])\chi_\textrm{m}[\omega]g\bar{a}(\delta F_\textrm{th}+\delta F_\textrm{tor}).\nonumber
\end{eqnarray}
Comparing Eqns.~(9) and (10) one can see that in both situations the measured output operator is proportional to the input quadrature operators and the sum of thermal force and the force applied via the magnetic torque mixing procedure described above. 

The above analysis leads to the conclusion that when operated in a homodyne optical circuit, the magneto-optomechanical transducer operates as a resonant optical phase modulator \cite{Ilchenko03} (\textit{e.g.},~electro-optic modulator (EOM)), whereas in direct detection it operates as an amplitude modulator.  In both situations the process is completely coherent such that the phase of the applied torque is directly transduced into the output optical signal.  Yet, as we experimentally demonstrate below, both of these optical signals are measured by the receiver as modulations to the mechanical phase, as expected from Eqn.~(7).  

This theoretical framework can be interpreted in the quantum regime by considering the driving magnetic field $H_z(f_z)$ to consist of single photons. In this limit, the transduced field would be dominated by thermal fluctuations at room temperature; however, by performing these experiments in a cryogenic environment the thermal force term can be eliminated \cite{Ramp18}, allowing the transduction of quantum signals.

\begin{figure}[t]
\includegraphics[width=0.45\textwidth]{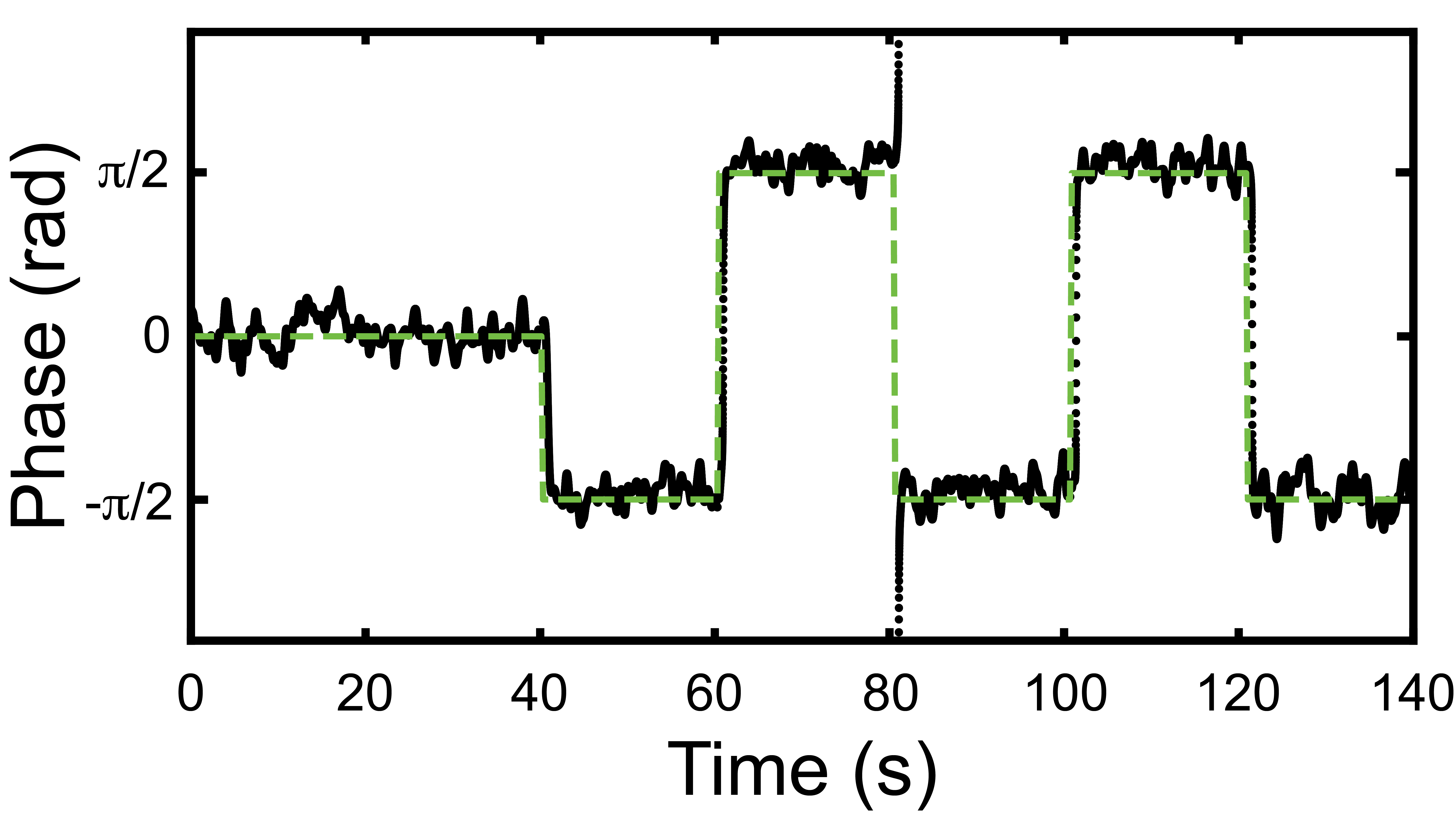}
\caption{Demonstration of the coherence of the demodulated 1570-nm-wavelength signal (black) with respect to the phase of the 401.7 MHz drive tone (green dashed).  }
\end{figure}

\section{Results and Discussion}
To test the coherent response of the wavelength conversion, we first use a homodyne optical detection system sensitive to shifts in the phase of the optical field within the silicon microdisk.  The microdisk lies in one arm of the homodyne system \cite{Hau18}, such that as the mechanical resonator moves in the evanescent field of the microdisk it shifts the optical phase in the cavity and hence the phase of the detected optical signal.  This optical signal is demodulated using a digital lock-in amplifier -- referenced to the same 10 MHz Rb source as the signal generators -- at the mechanical resonance frequency to attain the amplitude and, more importantly, phase of the signal modulated at $f_\textrm{m}$.  One can consider the photodetector and lock-in amplifier as the receiver in this carrier modulation scheme \cite{Anderson}.  The result, shown in Fig.~2, indicates that the phase of the demodulated optical signal at 1570 nm reliably follows the input phase of the UHF tone (here at 401.7 MHz).  

\begin{figure}[b]
\includegraphics[width=0.45\textwidth]{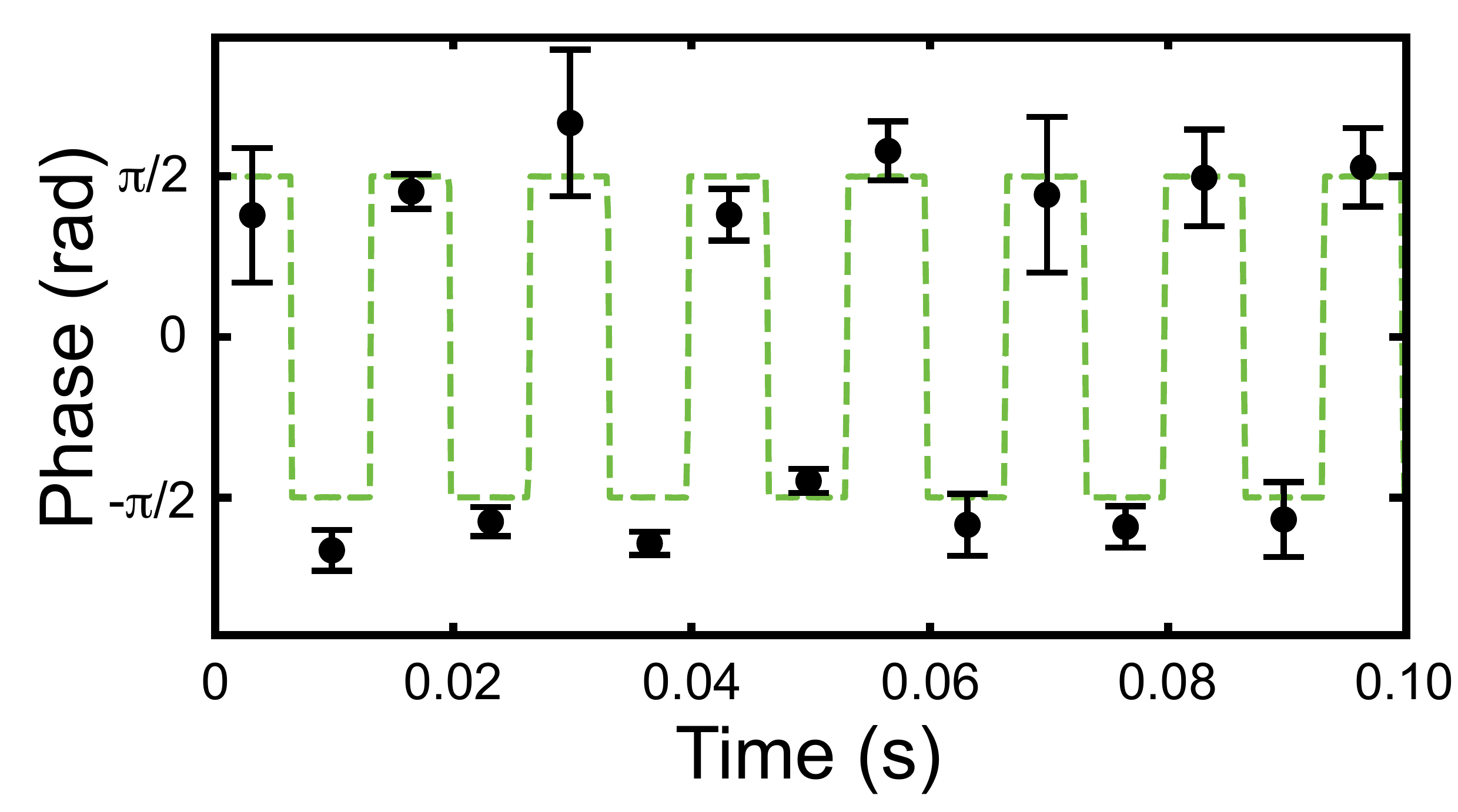}
\caption{Binary phase-shift keying encoded as ``non-return to zero'' \cite{Anderson} in the UHF tone (green dashed) is coherently observed in the phase of the telecom homodyne detection system.  Telecom phase data is averaged over 80\% of the time bin (of 6.7 ms), and assigned an average value (black circles) with error bars from the standard error of the mean.  This optomechanical wavelength conversion and data transmission rate corresponds to 150 baud. }
\end{figure}

\begin{figure}[b]
\includegraphics[width=0.45\textwidth]{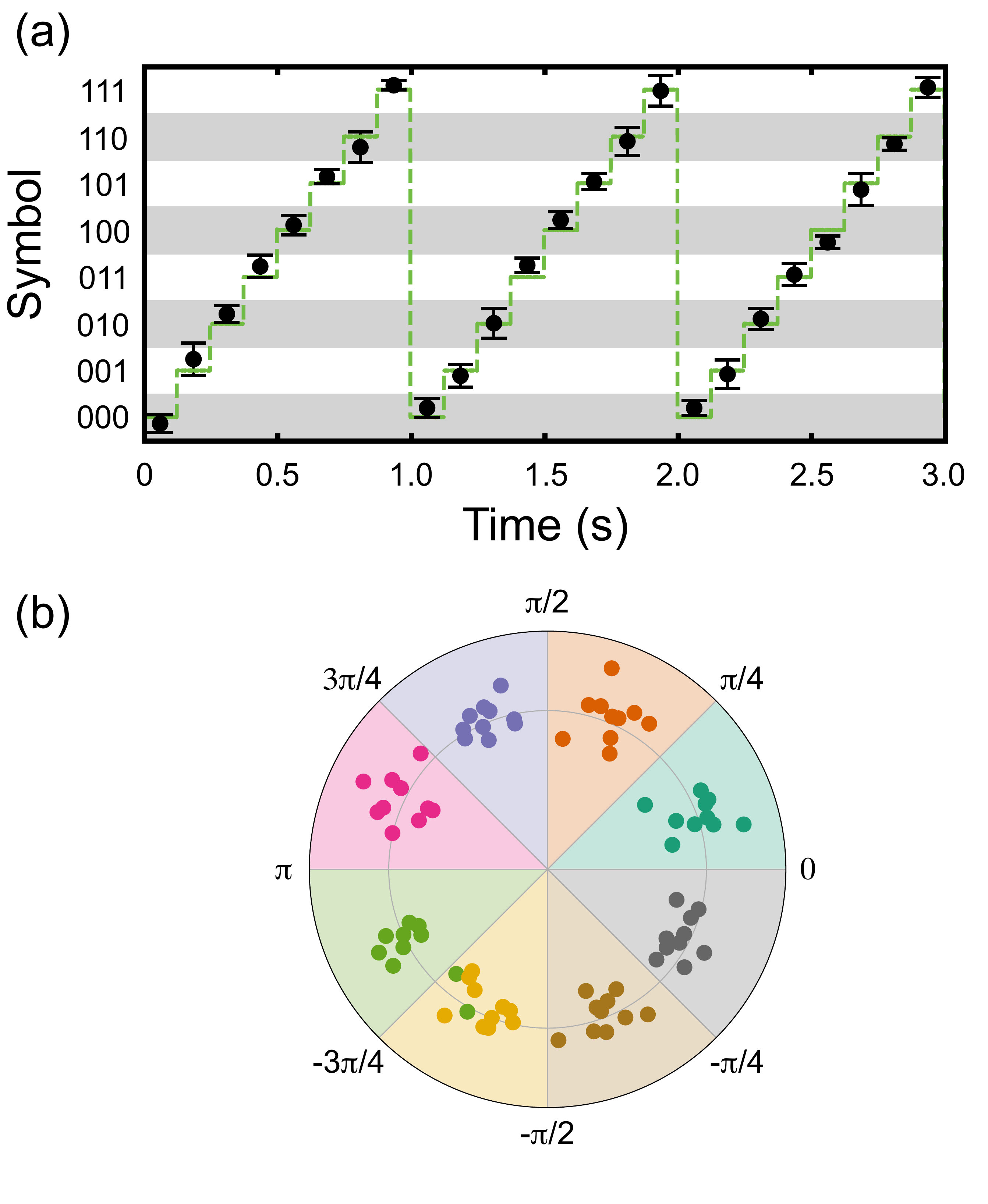}
\caption{a) Eight-bin phase-shift keying (8PSK), corresponding to 3 bits of information per time bin.  The phase of the 1570 nm telecom signal is averaged over 80\% of the time bin (black circles) and error bars correspond to the standard error of the mean. The phase of the 401.7 MHz tone is shown as the green dashed line. (b) 8PSK data represented on a constellation diagram for a larger data set.  Here, we demonstrate that 8PSK is the limit for high-order PSK with current levels of signal-to-noise in the demodulated phase.}
\end{figure}

For the purposes of transmitting information from a microwave source via a telecom carrier, one is naturally concerned with the speed at which the system can respond to changes in the phase.  To test this, we begin with binary phase-shift keying (BPSK), in which two phase bins -- corresponding to one bit -- are converted via torque mixing and read-out in the telecom fiber.   The signal is averaged over the middle 80\% of the time bin, resulting in an average phase with error bars associated with the standard error of the mean, shown in Fig.~3.  At a rate of 150 baud (symbols per second), we can associate each time bin with a particular phase bin.  This rate could be improved by increasing the signal-to-noise ratio of the converted signal, as discussed below.  

Interestingly, it is also possible to encode more than two phase bins in a single time bin, which has the potential to increase data transmission rates.  We show in Fig.~4 the phase of the microwave tones divided into 8 bins, or ``symbols,'' corresponding to 3 bits of data per time bin -- referred to as 8PSK \cite{Anderson}.   Unfortunately, this required slowing the transmission rate to 64 baud (3 bits in 8 symbols per 125 ms), due to the increased signal averaging needed to resolve all eight phase bins.  Therefore, in our current system higher-order PSK does not result in an improved data transmission rate, though with improvements in our signal-to-noise (so that individual phase bins are more easily resolved with less averaging) high-order PSK is likely to be advantageous because of the coherent response of the optical modulation to the phase of the input tone.  

\begin{figure}[t]
\includegraphics[width=0.4\textwidth]{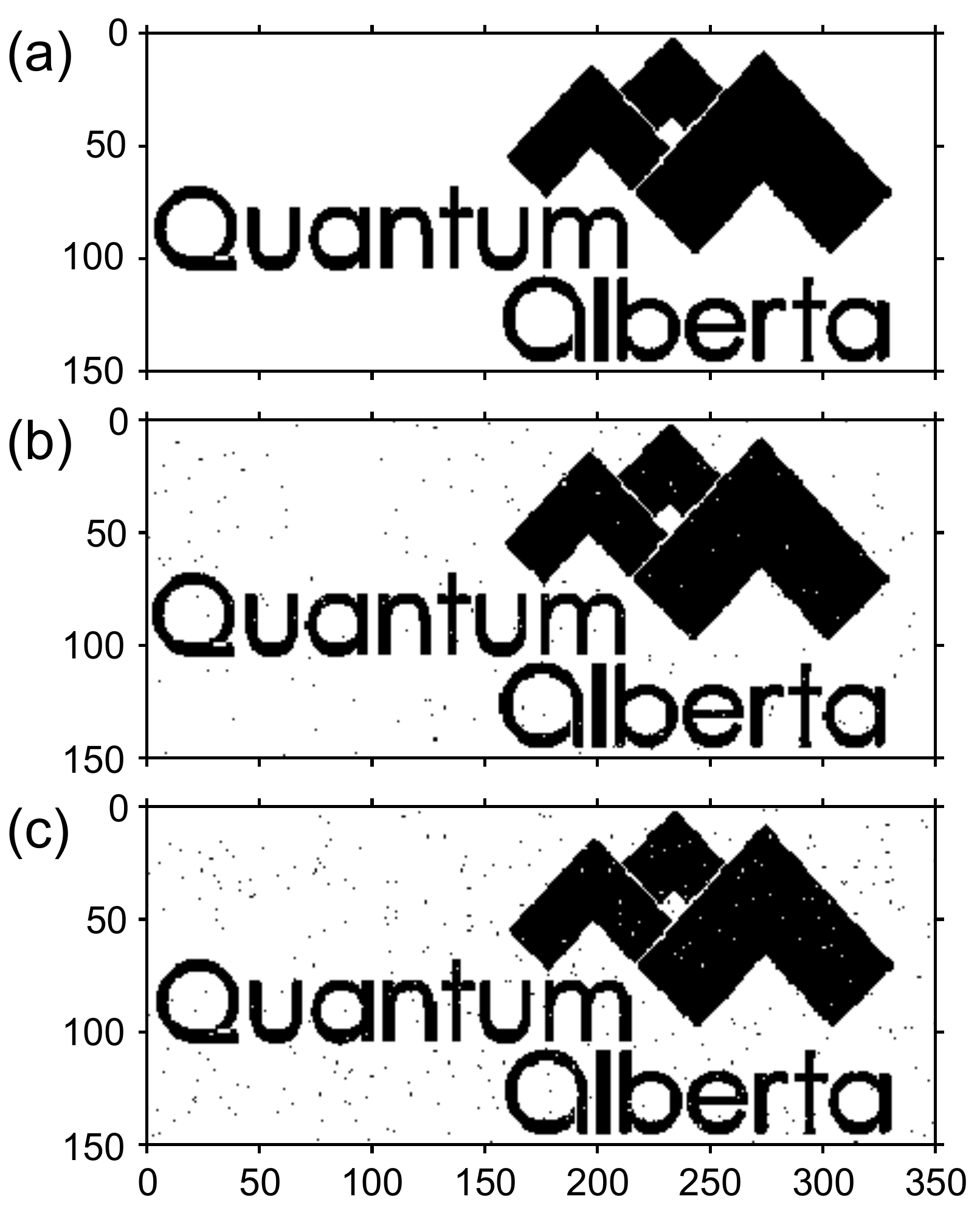}
\caption{a) Image consisting of $150\times350$ points (52,500 pixels).  Each pixel can be transmitted as a time bin, with the symbol (black or white) corresponding to the transmitted phase.  (b) Reconstructed image after being encoded in the UHF phase and demodulated after homodyne optical detection. Here, the error rate is 0.38\%.  (c) The image was encoded identically to panel b, but was transmitted via direct detection, as described in the text, allowing reconstruction after being transmitted down a 6 km long optical fiber.  Here, the error rate is 0.67\%.}
\end{figure}

Finally, after showing that our magneto-optomechanical wavelength conversion is coherent, we turn to demonstrating image transmission via BPSK.  As a proof of principle, we have constructed an image containing 52,500 pixels, chosen to be only black or white so that we have an equivalent number of bits of information to encode, as shown in Fig.~5a.  The phase of the microwave tone was then modulated to encode the bits, which were sent at a transmission rate of 150 baud to the out-of-plane drive coil.  The resulting telecom optical signal was demodulated as above, now with each time bin corresponding to one pixel, allowing reconstruction of the image as shown in Fig.~5b.  It is clear that this generates a faithful representation of the original image, with the introduction of just 0.38\% error in the transmitted pixels.

For networking applications it is impractical to imagine running two fiber optics for each information channel so that homodyne detection can be performed on a balanced photodetector.  Instead, we turn to the direct detection method described in Section III.  In this approach the amplitude of the optical signal is modulated coherently, but the phase of the mechanics is demodulated using the digital lock-in amplifier exactly as in the homodyne case.  In this configuration we can transmit the information encoded in the phase of the microwave tone over long distances in a single fiber optic cable, as shown for 6 km of fiber in Fig.~5c.  Here, the error rate increases to 0.67\%, which could be reduced by the introduction of a fiber-amplifier, although such amplification is not applicable to quantum information transmission.

\section{Conclusion}
We have demonstrated that magneto-optomechanics enables coherent signal transduction from UHF tones to telecom wavelengths.  We have used this coherence to demonstrate (classical) optomechanical information transduction via phase-shift keying, using both binary and eight-bin phase-shift keying.  Future improvements in signal-to-noise, for example by using lower spin-damping materials such as yttrium-iron-garnet \cite{Zhang14,Osada16,Harder17,Graf18}, could allow high-order phase-shift keying to be used in classical networking applications.  Significant improvements could also be made by improving the overlap between the magnetic field from the drive tone and the nanoscale magnetic component on the torsional resonator, for example, by using low-impedance microwave resonators \cite{Bienfait16,Eichler17,McKenzie19}.  Such low-mode volume superconducting resonators may also be vital for extending this technique to wavelength transduction in the quantum regime.  For quantum-level transduction, one would also need to eliminate thermomechanical noise, which is difficult in MHz-frequency devices \cite{Kim16,Hau18}.  To facilitate this, it may be possible to construct optomechanical crystal torque sensors \cite{Wu17} with higher mechanical frequencies that would more easily allow for ground-state preparation \cite{Ramp18,Forsch18}, or to use carefully timed dynamical back-action cooling as has been performed for microwave drum resonators \cite{Palomaki13}.  With such avenues for improvement, magneto-optomechanical signal transduction has the potential to be useful for quantum, as well as classical, information transmission.  

\begin{acknowledgments}
The authors thank L.J. LeBlanc for loaning her 6 km fiber spool.  This work was supported by the University of Alberta, Faculty of Science; the Natural Sciences and Engineering Research Council, Canada (Grants Nos. RGPIN-04523-16, DAS-492947-16, STPGP 494024-16, and CREATE-495446-17); and the Canada Foundation for Innovation. 
\end{acknowledgments}

\appendix

\section{Equations of motion}

In our experimental setup the optical resonator is driven by a coherent state, evanescently coupled via a dimpled tapered optical fiber \cite{Michael07,Hau14}.  The input coherent state is described by $a_\textrm{in} = (\bar{a}_\textrm{in} + \delta a_\textrm{in}(t))e^{i\omega_\textrm{l}t}$, where $\bar{a}_\textrm{in}$ is the coherent state amplitude, $\delta a_\textrm{in}$ are small fluctuations about the coherent state amplitude, and $\omega_\textrm{l}$ is the frequency of the driving laser. Furthermore, we can define the cavity field as $a = (\bar{a}+\delta a(t))e^{i\omega_\textrm{l}t}$, where $\bar{a} = \sqrt{\kappa_\textrm{e}}\bar{a}_\textrm{in}/(\kappa/2 - i\Delta_\textrm{l})$ is the intra-cavity coherent state amplitude, $\delta a(t)$ are small fluctuations about this steady state amplitude, and $\Delta_\textrm{l} = \omega_\textrm{l} - \omega_\textrm{c}$ is the laser detuning, where we have included any constant shift of the optical resonance due to radiation pressure. Using this definition, and assuming $\bar{a}$ is real, the value $\bar{a}^2 = \bar{N}$ is the average number of intra-cavity photons. We can linearize the equations of motion about this defined steady state, ignoring terms non-linear in the small fluctuation parameters and arriving at the equations,
\begin{eqnarray}
x&&[\omega] = \frac{\chi_\textrm{m}[\omega]}{1-i\hbar g^2\bar{a}^2\chi_\textrm{m}[\omega](\chi_\textrm{c}[\omega]-\chi^*_\textrm{c}[-\omega])} \\
&&\Big(\delta F_\textrm{th}+\delta F_\textrm{tor}-\hbar g\bar{a}\big(\chi_\textrm{c}[\omega]\delta a_\textrm{in}[\omega]+\chi^*_\textrm{c}[-\omega]\delta a^*_\textrm{in}[\omega]\big)\Big),\nonumber
\end{eqnarray}
\begin{equation}
    \delta a[\omega] = \chi_\textrm{c}[\omega] \Big(-ig\bar{a}x[\omega]+\sqrt{\kappa_e}\delta a_\textrm{in}[\omega]
    \Big).
\end{equation}
The term $i\hbar g^2\bar{a}^2\chi_\textrm{m}[\omega](\chi_\textrm{c}\textrm[\omega]-\chi^*_\textrm{c}[-\omega])$ modifies the mechanical susceptibility through the optical spring and optical damping effects for non-zero laser detuning. However, in the experimentally relevant regime when $C = 4\bar{N}g^2x_\textrm{ZPF}^2/\kappa \Gamma_\textrm{m} \ll 1$, where $x_{\rm zpf} = \sqrt{\hbar / 2 m \omega_{\rm m}}$ is the mechanical zero-point fluctuation amplitude, these terms are small and may be ignored. Furthermore, the additional forcing term $\hbar g\bar{a}\big(\chi_\textrm{c}[\omega]\delta a_\textrm{in}[\omega]+\chi^*_\textrm{c}[-\omega]\delta a^*_\textrm{in}[\omega]\big)$ is the optical backaction resulting from the random radiation-pressure shot noise. This term is much smaller than the other forces in this experiment and may also be ignored. However, if the force resulting from the torque mixing process is quantum in nature, this term would limit the quantum efficiency of the transduction to the standard quantum limit. Ignoring these terms we are left with the simplified equation given in the main text. 

\section{Optical noise quadratures}

The input optical amplitude and phase quadrature operators are given by  \cite{Loudon00},
\begin{equation}
\delta X_{I,\textrm{in}} = \delta a_\textrm{in} + \delta a^*_\textrm{in},
\end{equation}
\begin{equation}
\delta X_{Q,\textrm{in}} = -i(\delta a_\textrm{in} - \delta a^*_\textrm{in}).
\end{equation}
Using the input-output formalism that states $a_\textrm{out} = a_\textrm{in} - \sqrt{\kappa_e}a$, where the output field may be written as $a_\textrm{out} = (\bar{a}_\textrm{out} + \delta a_\textrm{out}(t))e^{i\omega_\textrm{l}t}$, one can solve for the output optical quadrature operators. Here, $\bar{a}_\textrm{out}$ is the coherent output state amplitude, $\delta a_\textrm{out}$ are small fluctuations about the coherent output state amplitude. If we consider the noise spectrum of the output field we can define the output amplitude and phase quadrature operators as,
 \begin{equation}
     \delta X_{I,\textrm{out}} = \delta a_\textrm{out} + \delta a^*_\textrm{out},
 \end{equation}
 \begin{equation}
     \delta X_{Q,\textrm{out}} = -i(\delta a_\textrm{out} - \delta a^*_\textrm{out}).
 \end{equation}
These may be solved for directly using Eqn.~(8), and the relationship between the input and output field defined above to obtain Eqn.~(9) and (10) in the main text.

\end{document}